\def\beq{\begin{equation}}

\newif\iffigs\figsfalse

\figstrue

\iffigs
\input epsf.tex

\else

\fi

\def\bbb1{{\rm 1\!1}}

\newcommand{\refs}[1]{(\ref{#1})}

\documentclass[12pt]{JHEP}

\usepackage{epsfig}

\advance\hoffset by -.35in
\def\>{\rangle}
\def\<{\langle}

\def\al{\alpha}

\def\al{\alpha}

\preprint{KCL-TH-01-09}

\title{\Large\bf On Higher Derivative 
Terms in Tachyon Effective Actions}

\author{N.D. Lambert${}^a$ and I. Sachs${}^b$\\

\vskip 24pt

$^{a}$Dept. of Mathematics\\
King's College\\
The Strand\\
London, WC2R 2LS\\
England\\

\vskip 24pt

$^{b}$Theoretische Physik\\
Ludwig-Maximilians Universit\"at\\
Theresienstrasse 37 \\80333 Munich\\ Germany }

\abstract{
We reconstruct the tachyon effective action for unstable 
D-branes in superstring theory by examining its behaviour near 
exactly marginal 
deformations, where the ambigous higher derivative terms can be eliminated. 
We then compare this action with that obtained in boundary string field 
theory and find remarkable agreement. In particular, the tension for lower 
dimensional branes and the BI-action for the centre of mass motion 
are reprodued exactly. We also comment on the action for tachyons on the kink 
in a D-brane/anti-D-brane system and on bosonic string theory. }

\begin{document}

\section{Introduction}
Non-BPS branes play an important role in the non-perturbative dynamics of 
string theory \cite{9701137}-\cite{9911081} as well as 
non-supersymmetric field theory \cite{9802183}-\cite{0010045}. 
Contrary to BPS branes \cite{9510017}, 
which correspond to stable supersymmetric states, non-BPS branes are 
generically  unstable and decay into the 
closed string vacuum or a BPS/anti-BPS brane pair. 
This instability manifests itself by a tachyonic mode 
in the open string sector \cite{9511194}. The condensation of this tachyon 
into a stable 
vacuum is a very interesting though challenging off-shell problem in string 
theory (see e.g. \cite{Halpern} for early discussions on tachyon 
condensation in bosonic string theory). In particular, unlike the Born-Infeld (BI) action of 
the massless degrees of freedom on a BPS-brane 
\cite{Fradkin85,Tseytlin86,Callan87}, obtaining an effective action for the 
tachyon is not straightforward \cite{0008231,0009246,0011226}. 
Nevertheless, over the last couple of years considerable 
progress has been made, both within string field theory (SFT) 
\cite{Witten86} and the 
$\sigma$-model approach (BSFT) \cite{9208027,9303143,9311177}. 
Indeed Sen's relation \cite{9805170} between brane tension and 
the tachyon potential energy has been acurately verified  using 
level truncation in SFT \cite{Samuel}-\cite{0005036}. On the other hand, 
a portion of the tachyon effective action 
was probed by perturbing the open string $\sigma$-model with specific 
tachyon profiles \cite{0003101}-\cite{Takayanagi}. In particular this led 
to the exact tachyon 
potential by probing the theory with a constant profile. 
Linear, or quadratic profiles can also be 
treated in this way. However, extracting an 
effective action from 
the corresponding partition function is ambigous due to the higher derivative 
terms that vanish on this particular profile \cite{0011009}. 
For example terms in the 
Lagrangian of the form $T\partial^2 T$ vanish on a linear profile but, 
after integration by parts, contribute to the standard kinetic term.
Nevertheless the  tensions obtained for various 
solitons  exactely match the tensions of the lower dimensional 
D-branes they are supposed to describe. 

In this paper we reconsider the issue of higher derivative terms in the 
tachyon 
effective action. Our approach is based on combining relevant and marginal 
perturbations of the open string $\sigma$-model. Specifically we probe the 
effective action with an  exactely marginal tachyon profile. This has the 
advantage that, on this particular background, all higher derivatives can be 
eliminated. We can therefore construct an effective action with only 
first order
derivatives by imposing that the equations of motion are satisfied exactly 
on this profile. As we shall 
see, the action obtained in this way is then determined uniquely in terms of 
a potential function. The potential  cannot 
be determined without extra information. 

One way to proceed is simply to 
take the exact tachyon potential from BSFT. This is equivalent 
to neglecting higher derivative terms, since the extra terms in the potential 
generated by eliminating higher derivative terms on the marginal profile 
are not taken into account. Nevertheless we ensure 
that the exact conformal background corresponding to the marginal deformation 
is always an extremum of the truncated action. Of course the elimination of 
higher derivative terms works only on this specific profile. Hence one might
only expect that  
this action is a good approximation of the tachyon dynamics only around 
the marginal profile. However, we find 
that that our action agrees 
remarkably well with that obtained in BSFT constructed from a very 
different profile, i.e. from a linear tachyon. In fact, 
it matches the latter exactly for linear profiles with large slope $T'$, but 
in contrast to the latter reproduces the correct perturbative mass of the 
tachyon by construction. This form also
produces the expected tension of the lower dimensional branes. 
Furthermore, we show that our action leads to the expected BI-action for 
the centre of mass fluctuation of the lower-dimensional branes 
\cite{0006122,Theisen}. That  the two actions are almost 
identical comes as a surprise and one may 
be tempted to take this as a hint that, rather like for BPS states, higher 
derivative corrections are unimportant on marignal profiles. However, 
we know of no fundamental principle leading to such 
behaviour (for example 
note that higher derivatives cannot be eliminated for linear 
profiles).  

We can also apply this proceedure to $D\bar D$-systems in superstring theory. 
The remarkable aggreement between our action and the BSFT action contunies 
to hold in this case. A new feature 
for the $D\bar D$-system is the existence of a tachyonic mode in addition to 
the massless centre of mass zero mode. We find a Born-Infeld type action for 
the tachyonic mode on a linear profile as suggested in \cite{Garousi,0003221} 
but the kinetic term is,  
however, scaled to zero by the renormalisation group flow. 

One can also apply our proceedure to the bosonic string. In this 
case we have not been able to find a closed form for the effective action.
In addition there is no closed form available from BSFT and therefore
it is difficult to make many comparisons  
\cite{0009103,0009148}. However we do find agreement for the lowest 
order tachyon kinetic terms. 

An alternative way to approximate the potential function is to impose that 
the tachyon 
action should reproduce the BI-action for the massless centre of mass 
fluctuations of stablised non-BPS brane wrapped on an orbifold
\cite{0006122,Theisen}. We find that this fixes the  action uniquely and 
leads to an action in qualitative agreement with that found by the other 
methods, although again we have been unable to find a closed expression
for the various coefficients and hence our analysis is incomplete. 

The rest of this paper is organised as follows. In the next section
we discuss some general features of kink solitons in arbitrary scalar
Lagrangians that are first order in derivatives.  We then proceed to
construct particular Lagrangians relevant to tachyons on a 
non-BPS brane by demanding that a class of exactly marginal deformations
solve the equations of motion. This fixes the Lagrangian  in terms
of the potential. In the following three sections we then proceed 
by inputing the potential found in BSFT for the cases of a non-BPS
D-brane, a D-brane/anti-D-brane pair and D-branes in Bosonic string
theory. In section four we propose an alternative derivation of 
the tachyon Lagrangian by demanding that it reproduce the
effective action of the massless fields for a stablised non-BPS 
branes on an orbifold. Section five contains a brief summary of our results.

\section{The Tachyon Action}

\subsection{General Comments on Kinks}

Consider a non-BPS D$p$-brane whose worldvolume
spanned by the coordinates $x^m$, 
$m=0,1,2,3,..,p$. 
The most general form for the effective
action of a real $T$ in $p+1$ dimensions that
depends on at most first order derivatives and is even in $T$ is 
given by
\begin{eqnarray}\label{tg1}
S &=& \int d^{p+1} x {\cal L} 
\equiv \int d^{p+1} x \sum_{\alpha, \beta=0}^{\infty} c_{\alpha\beta}T^{2\alpha}
(\partial_m T\partial^m T)^\beta\ .
\label{action}
\end{eqnarray}
Consider now the set of static solutions for $T$ depending on 
only one space-like coordinate, say $x^p$, 
and denote $T' = \partial_{p} T$.
The resulting equation of motion is
\begin{eqnarray}
 \sum_{\alpha, \beta=0}^{\infty}\left[ 2\alpha(1-2\beta) 
c_{\alpha\beta}T^{2\alpha-1}(T')^{2\beta} 
-2\beta(2\beta-1) c_{\alpha\beta}T^{2\alpha}(T')^{2\beta-2}T'' 
\right]=0\ .
\label{eqofm1}
\end{eqnarray}
For a non-trivial kink solution we can multiply \refs{eqofm1} by $T'$ and
find
\begin{eqnarray}
 \sum_{\alpha, \beta=0}^{\infty}\left[
(1-2\beta)c_{\alpha\beta}T^{2\alpha}(T')^{2\beta} 
\right]'=0\ .
\label{eqofm2}
\end{eqnarray}
Thus we arrive at the first order equation for a kink solution
\begin{equation}
{\cal L} - {\delta {\cal L}\over \delta T'}T' = V_0\ ,
\label{kinkeq}
\end{equation}
that is, the Legendre transform
of the Lagrangian evaluated on a Kink solution is  constant. 
This is, of course, a general result valid for any Lagrangian depending only on 
$T$ and it first derivative\footnote{Note that \refs{kinkeq} is also 
fulfilled for radially symmetric tachyon solutions $T(r)$.}. 
As an example we can consider the class of actions of the form
\begin{equation}
{\cal L} = V(T)K(\partial_m T\partial^m T)\ .
\label{ex1}
\end{equation}
In particular,  
the tachyon effective action proposed in \cite{Garousi} 
is of this form with
$K = \sqrt{1+\partial_m T\partial^m T}$. However, the kink
equation becomes
\begin{equation}
V(T)K((T')^2) - V(T){d K((T')^2)\over dT'}T'= V_0\ .
\label{ex2}
\end{equation}
If the kink approaches the closed string vacuum at infinity, then, according 
to Sen's conjecture, $V_0=0$. In this case we see that the factors of 
$V(T)$ in \refs{ex2}
can be cancelled off and hence  the only possible kink solutions 
linear functions with $T'$ constant.
Moreover 
in the case that $K((T')^2)=\sqrt{1+(T')^2}$, the only solution is 
$T'=\infty$, as previously observed in \cite{0011226}.

\subsection{Constructing The Kink Action}

As is well known, when compactified on a circle of radius $\sqrt{2\alpha'}$ 
$T = \chi {\rm sin}(x/\sqrt{2\alpha'})$ 
is  an exactly marginal deformation 
to all orders in $\alpha'$ and for any constant $\chi$
(e.g. \cite{9812031,0006122}). We therefore demand that $T$ solves 
the field equations of the exact tachyon effective action. Furthermore because 
$T''=\frac{1}{2\al'}T$ we can eliminate all higher derivatives. That is, 
without restricting the generality we can assume that the the tachyon action 
on this profile has only first derivatives, i.e. is of the form \refs{tg1}. 

To continue it is helpful to replace the 
coefficients $c_{\alpha\beta}$ in the Lagrangian by 
\begin{equation}
c_{\alpha\beta} = \kappa^{\beta}
\left({\kappa\over 2\alpha'}\right)^\alpha {\tilde c}_{\alpha\beta}\ ,
\end{equation}
where $\kappa$ is a free parameter, that can be absorbed into a
redefinition of $T$, but which we keep for generality.
The equation of motion is given in \refs{eqofm2} and hence we find
\begin{eqnarray}\label{sum}
0=\sum_{\alpha,\beta}(1-2\beta){\tilde c}_{\alpha\beta}
\left({\kappa\chi^2\over 2\alpha'}\right)^{\alpha+\beta}
{\rm sin}^{2\alpha-1}\left({x\over\sqrt{2\alpha'}}\right)
{\rm cos}^{2\beta-1}\left({x\over\sqrt{2\alpha'}}\right)&&\nonumber\\
\cdot\left[2\alpha{\rm cos}^{2}\left({x\over\sqrt{2\alpha'}}\right)-2\beta{\rm sin}^{2}\left({x\over\sqrt{2\alpha'}}\right)\right]&& .
\end{eqnarray}
However, since we demand that \refs{sum} is true for all $\chi$ we may 
replace the sum over $\al,\beta$ by separate sums labelled by $n=0,1,2,...$ 
where $\alpha+\beta=n$. Hence for
each $n$ \refs{sum} yields $n$ linear conditions on the
coefficients ${\tilde c}_{\gamma\ n-\gamma}$. In particular we
find, for $\gamma = 1,...,n$,
\begin{equation}
{\tilde c}_{n -\gamma+1\ \gamma-1} = 
{\gamma(1-2\gamma)\over (3-2\gamma)(n+1-\gamma)}
{\tilde c}_{n -\gamma\ \gamma}\ .
\end{equation}
This in turn implies that all the coefficients ${\tilde c}_{\alpha
  \beta}$
with $\alpha+\beta=n$ can be determined in terms of ${\tilde c}_{n
  0}$ to be
\begin{equation}
{\tilde c}_{n -\gamma\ \gamma} = 
-{1\over 2\gamma-1}\left(\matrix{n\cr \gamma\cr}\right)
{\tilde c}_{n 0}\ .
\end{equation}
On the other hand the coefficients ${\tilde c}_{n 0}$ correspond to
the Taylor series expansion of the potential $V(T)$ in powers of
$\kappa T^2/{2\alpha'}$. Thus the requirement that 
$T = \chi {\rm sin}(x/\sqrt{2\alpha'})$ solves the field equations 
completely determines the action
\refs{action} in terms of an arbitrary potential $V(T)$.
We may further clarify the situation by constructing the Lagrangian 
explicitly in terms of $V(T)$
\begin{equation}\label{KV}
{\cal L} = -\sum_{\gamma=0}^{\infty}{1\over \gamma !}{1\over 2\gamma-1}
{d^\gamma V(t)\over dt^\gamma}(\kappa \partial_mT\partial^mT)^\gamma\ ,
\end{equation}
where $t\equiv \kappa T^2/{2\alpha'}$.
This is a highly suggestive form for the Lagrangian. It's meaning
is easily obtained by evaluating the resulting kink equation \refs{kinkeq}
\begin{eqnarray}
V_0={\cal L} - {{\cal L}\over d T'}T'
&=& \sum_{\gamma=0}^{\infty}{1\over \gamma !}
{d^\gamma V(t)\over dt^\gamma}(\kappa (T')^2)^\gamma\cr
&=& V\left({\kappa T^2\over{2\alpha'}} +\kappa(T')^2\right)\ .
\label{kinkagain}
\end{eqnarray}
Thus, assuming that the minima of $V$ are isolated, we see that
the only regular solutions are
\begin{equation} 
T = \chi {\rm sin}\left({x-x_0\over\sqrt{2\alpha'}}\right)\ ,
\label{tkink}
\end{equation}
for arbitrary $x_0$ and $\chi$.  
To continue we need to fix the function $V(T)$. There is no direct way to 
construct $V(T)$ exactly in this simple set-up. 
In the next  subsections 
we explore the results obtained by taking the potential from BSFT  
and compare them with previous results for the cases of a non-BPS D-brane,
D-brane/anti-D-brane pairs and D-branes in bosonic string theory.

\subsection{Matching BSFT: non-BPS D-Branes}

In boundary string field theory (BSFT) the exact 
tachyon potential on a non-BPS D-brane can be determined by computing 
the disk partition function on a constant tachyon profile  
and yields \cite{0010108,0012198}
\begin{equation}\label{VBSFT}
V(T) = \sqrt{2}\tau_pe^{-{\kappa T^2\over 2\alpha'}}\ ,
\end{equation}
where $\tau_p$ is the tension of a BPS $Dp$-brane. Of course, the elimination 
of higher derivative terms on our kink background will affect the 
non-derivative terms and hence modify the potential in a, a priori 
uncontrolled way. 
In this section we will simply ignore these terms, that is, we substitute  
\refs{VBSFT} in \refs{KV}. Doing so is equivalent to ignoring certain 
higher derivative terms. The resulting  tachyon action nevertheless has some 
desirable features by construction. In particular, it will admit the exact 
kink \refs{tkink} as a solution and 
the perturbative tachyon mass is also reproduced correctly. 
The resulting Lagrangian that we construct then takes the
form
\begin{eqnarray}\label{kinkyaction}
{\cal L} &=& -\sqrt{2}\tau_pe^{-{\kappa T^2\over 2\alpha'}}
\sum_{\gamma=0}^{\infty}{1\over \gamma!}{1\over 2\gamma-1}(-\kappa
\partial_m T\partial^m T)^\gamma \\
&=& \sqrt{2}\tau_pe^{-{\kappa T^2\over 2\alpha'}}
\left[e^{ -\kappa\partial_m T\partial^m T}
+ \sqrt{\pi\kappa\partial_m T\partial^m T}
{\rm erf}\left(\sqrt{\kappa\partial_m T\partial^m T}\right)
\right] 
\ ,\nonumber 
\end{eqnarray}
where 
\begin{equation}
{\rm erf}(x) \equiv {2\over\sqrt{\pi}}\int_0^x e^{-t^2}dt \ .
\end{equation}
Let us now compare the action found here to that obtained in 
BSFT. In  \cite{0010108,0012198} the effective
action on a non-BPS D-brane was obtained to all orders in 
$T$ and $\partial_m T\partial^m T$ but ignoring other higher
derivative terms (more accurately the tachyon was assumed
to have a linear form so that only these derivative terms
are non-vanishing). Their result was 
\begin{equation}
{\cal L} = \sqrt{2}\tau_p
e^{-{T^2\over 2\alpha'}}{1\over2}{4^y y\Gamma(y)^2\over\Gamma(2y)}\ ,
\label{bsftaction}
\end{equation}
where $y = \partial_mT \partial^m T$.\footnote{The action
in \cite{0012198} seems to differ by $y\rightarrow y/2$.} 
Thus the overall potential factor appears identically in both
\refs{kinkyaction} and \refs{bsftaction}. By comparing the potentials we
see that $\kappa=1$. Let us then compare
the kinetic terms. These have been plotted in figure one, i.e.  
${1\over2}{4^y y\Gamma(y)^2/\Gamma(2y)}$ and 
$e^{-y}+\sqrt{\pi y}{\rm erf}(\sqrt{y})$ are plotted 
as a function of $y$.  Although the power expansions
of the two functions are different, there is a remarkable agreement
between the two. A notable difference is that, by construction, the effective
action \refs{kinkyaction} produces the correct perturbative tachyon mass 
($m^2=-1/2\alpha'$), whereas in BSFT expression \refs{bsftaction} 
one finds $m^2 =-1/4\alpha'{\rm ln}2 $. The difference is related a 
different treatment of the second derivatives in the two approaches. 

In addition to the marginal deformations discussed so far, the linear 
tachyon profile, $T=ux$ is important. It corresponds to the IR fixed point 
(for $u=\infty$) of the renormalisation group flow of relevant 
perturbations and describes a D$(p-1)$-brane \cite{0010108}. 
It is clear that $T=ux$ where $x$ is one of the coordinates, say $x^p$, 
of the non-BPS $Dp$-brane,
cannot be compatible with \refs{kinkagain}. However, if we allow for singular 
configurations then $T=ux$ is a solution
if we scale $u\rightarrow \infty$. Indeed we can consider any function 
of the form $T(x) = u\tilde T(x)$ with $u\rightarrow\infty$, provided
that $\tilde T(x)=0$ only on discrete set of points.
The energy of such a  configuration is
\begin{eqnarray}
E &=& \sqrt{2} \tau_p \int_{-\infty}^{+\infty}
e^{-{\kappa u^2\over2\alpha'}\tilde T^2}
\left(e^{-\kappa u^2({\tilde T}')^2}+\sqrt{\pi\kappa}u|\tilde T'|
{\rm erf}(\sqrt{\kappa}u|\tilde T'|)
\right)dx \cr
&=& \sqrt{2\pi\kappa}u\tau_p\int dx\, |\tilde T'| 
e^{-{\kappa u^2\over2\alpha'}\tilde T^2}\cr
&=&2\pi N\sqrt{\alpha'}\tau_p\ ,
\end{eqnarray}
where in the second line we took the limit $u\rightarrow\infty$.
Here $N$ is the number of times $T(x)$ covers the real line, which
in the $u\rightarrow\infty$ limit is simply the number of times
$\tilde T$ changes sign.
This is precisely the correct value to interpret the kink as
$N$ $D(p-1)$-branes. 
Note that in this limit it does not matter if
we instead use the kinetic terms given in 
\refs{bsftaction}
since they have the same asymptotic value for large $u$. 
This provides some support for the claim that polynomial 
kinks of the form $T= u x^N +...$ correspond to $N$
multiple lower dimensional 
branes in the limit $u\rightarrow\infty$  \cite{0003101}. 

Next we may derive the low energy dynamics on multi-kink 
$T=u\prod_{i=1}^N(x-a_i)$
by letting the zero modes $a_i$ depend on the other 
coordinates $x^\mu$, $\mu=0,...,p-1$ of the non-BPS $Dp$-brane.
Substituting in to the action \refs{kinkyaction} we find, taking the
limit $u\rightarrow\infty$,
\begin{equation}
{\cal L}_{Kink} = \sqrt{2\kappa\pi}\tau_p u\int dx\,|\tilde T'|\, 
e^{-{\kappa u^2\over 2\alpha'}\tilde T^2}
\sqrt{1+{\partial_\mu \tilde T\partial^\mu \tilde T\over (\tilde T')^2}}
\ .
\label{kinkeffact}
\end{equation}
Next we note that
\begin{equation}
{\partial_\mu \tilde T\partial^\mu \tilde T\over (\tilde T')^2}
 = \sum_{i,j}{\partial_\mu a_i\partial^\mu a_j\over (x-a_i)(x-a_j)}
/\left(\sum_{k}{1\over (x-a_k)}\right)^2 \ .
\label{hhh}
\end{equation}
We assume that all the $a_i$ are distinct. 
On each branch  one can invert $\tilde T(x)$ to find 
$x(\tilde T)=x(u^{-1}T)$  and substitute
\refs{hhh} into \refs{kinkeffact}. However 
we need only consider the large $u$ limit, 
which is equivalent to considering $x(0)=a_i$. The result is 
the integral \refs{kinkeffact} includes a  sum 
over the $N$ zeros $x=a_i$ and leads to
\begin{equation}
{\cal L}_{Kink} = 2\pi\sqrt{\alpha'}\tau_p\sum_{i=1}^N
\sqrt{1+\partial_\mu a_i\partial^\mu a_i}\ .
\end{equation}
This is precisely the Born-Infeld action for the massless modes
of $N$ BPS $D(p-1)$-branes where the 
$a_i$ parameterise their separations.

\subsection{$Dp$-brane/${\bar D}p$-brane pairs}

It is clear that 
a similar action is constructed for a $Dp$-brane/${\bar D}p$-brane pair,
only with $T$ complex so that  $T^2$ replaced by $|T|^2$ 
and in addition there is an extra factor of $\sqrt{2}$ in the tension
\cite{0010108,0012198}. In these actions the same
linear kink solution should now represent
a non-BPS $D(p-1)$-brane. However there is now an additional tachyonic
mode since if we consider fluctuations we must
set $T = u(x-x_0)+ it$, where $t$ is real and hence cannot be absorbed
into $x_0$ (here, for simplicity, we consider only a single kink).
With
this ansatz the effective action for the kink is, in the large $u$
limit, 
\begin{eqnarray}
\cal L &=& 2\tau_p\int dx e^{-{\kappa u^2\over 2\alpha'}(x-x_0)^2}
e^{-{\kappa^2t^2\over 2\alpha'}} \sqrt{\kappa} u
\sqrt{1+\partial_\mu x_0\partial^\mu x_0+u^{-2}\partial_\mu t\partial^\mu t}
\nonumber\\
&=& \sqrt{2}\tau_{p-1}\,e^{-{\kappa^2t^2\over 2\alpha'}}
\sqrt{1+\partial_\mu x_0\partial^\mu x_0}\ .\nonumber\\ 
\end{eqnarray}
Note that the kinetic term for the remaining tachyon mode $t$ 
has disappeared. Had we rescaled $t\rightarrow ut$ in order to
obtain a finite kinetic term for $t$, the entire action would have vanished
due to the exponential factor. Apart from this issue we arrive at
the Born-Infeld effective action for a non-BPS $D(p-1)$-brane.

\subsection{The Bosonic String}

In principle the application of our approach to the bosonic string 
is straightforward.  The only difference to the superstring is that 
the potential is no longer an even function of $T$.
The tachyon Lagrangian is then found to have 
a similar form, so that the coefficients $c_{\alpha\beta}$ 
are again be determined 
by a recursion relation in terms of the coefficients of potential
in a Taylor expansion. We find
\begin{equation}
{\cal L}=-\sum\limits_\beta\frac{1}{(2\beta-1)}\sum\limits_\alpha
\frac{(\alpha+2\beta)!!}{\alpha!!(2\beta)!!}c_{\alpha+2\beta\ 0}
T^\alpha
(\alpha'\partial_m T\partial^m T)^\beta\ ,
\end{equation}
where $n!!$ is the product over all positive odd (even) integers
that are less than or equal to $n$ if $n$ is odd (even). Unfortunately
there does not seem to be a simple formula analogous to \refs{KV} in 
this case.

In BSFT the potential is $V(T) = (1+T)e^{-T}$ \cite{0009103,0009148}. 
We have not been able to find a closed form expression for the resulting 
Lagrangian ${\cal L}$. However the first few kinetic terms can be
readily found to be
\begin{equation}
{\cal L}=(1+T)e^{-T}+{1\over2}e^{-T}\partial_m T\partial^m T
+{1\over24 T}(1-e^{-T})(\partial_m T\partial^m T)^2+...\ .
\end{equation}
The first order kinetic term is in agreement with BSFT \cite{0009103} 
(in our convention the tachyon mass is $-1$). 
One might also try to calculate the energy of a kink, which is now
of the form $T=ux^2$ with $u\rightarrow\infty$. In a term by term 
expansion of the integral we find some divergences. However it is
not clear if the actual energy diverges or if this is merely and
artifact of the expansion. 

\section{Matching the orbifold BI-action} 

In this section we will discuss a different procedure which determines the 
the higher derivative terms and also the
potential function $V(T)$ in \refs{kinkagain}. It consists of comparing 
the tachyon effective action \refs{KV} with that of a stable non-BPS 
D$p$-brane on a orbifold \cite{0002061,0006122}. Stable non-BPS 
branes can be obtained by 
wrapping a non-BPS D-brane over the orbifold ${\bf T}^4/{\bf Z}_2$.
Only  tachyon modes with odd units of momentum the orbifold directions 
survive. Thus the lightest modes of the tachyon have the form
\begin{equation}\label{k3}
T(x) = \sum_{i=6}^9 \chi^i(x^\mu)\ {\rm sin}\left({x^i\over
    R^i}\right)\ .
\end{equation}
In particular, at the critical radius $R^i = \sqrt{2\alpha'}$,
the momentum modes $\chi^i$ are exactly massless to all orders
in $\alpha'$. The 
effective action for $\chi^i$ was derived in \cite{0002061,0006122}. 
This action for a stabilised non-BPS brane
has a non-Abelian Dirac-Born-Infeld form, even for a single brane where  
the gauge group is $U(1)$ \cite{0006122}.
However, in the special case with non-vanishing momentum in only one 
orbifold direction, say $\chi^9$, the action of  \cite{0006122} 
is exact, up to higher 
derivative terms. For our purpose, that is 
to determine the effective action for an unstable non-BPS brane, 
this configuration is sufficient. 
The action is then simply
\begin{equation}
\mu \int d^{p-3}x
\sqrt{1+\kappa'\partial_\mu\chi^9\partial^\mu\chi^9}\ ,
\label{stableaction}
\end{equation}
up to higher derivative terms. 
Here $\mu$ are $\kappa'$ are constants which we have introduced for
the sake of generality. 

One might try to take the action \refs{kinkyaction} 
obtained in the previous section
and wrap it over the orbifold to find the effective action for the
mode $\chi^9$. This leads to 
\begin{eqnarray}
{\cal L} = -\sqrt{2}V\tau_p&&\sum_{\gamma=0}^\infty
\sum_{\delta=0}^\infty\sum_{k=0}^\gamma{1\over 2\gamma-1}
{1\over \delta!}{1\over k!}{1\over(\gamma-k)!}{1\over(\delta+\gamma)!}
\nonumber\\
&&\ \ \ \ \ \ \ 
{\Gamma(\gamma-k+{1\over2})\Gamma(\delta+k+{1\over2})
\over (\Gamma({1\over2}))^2}
\left(-{\kappa \chi_9^2\over 2\alpha'}\right)^{\delta+\gamma-k}
\left(-\kappa\partial_\mu\chi^9\partial^\mu\chi^9\right)^k\ ,\nonumber\\
\label{notsogood}
\end{eqnarray}
where $V$ 
is the volume of the orbifold at the critical radius.
This form is rather complicated and clearly differs from the 
expression \refs{stableaction}. Although the potential in \refs{notsogood}
(i.e. the $k=0$ terms) vanishes, one finds that the
kinetic terms depend on undifferentiated $\chi^9$'s. However, at
$\chi^9=0$, one finds 
\begin{equation}
{\cal L} = {V\tau_p\over \sqrt{2}}\sum_{\gamma=0}^\infty
\left(\matrix{{1\over2}\cr\gamma\cr}\right){1\over \gamma!}
(\kappa\partial_\mu\chi^9\partial^\mu \chi^9)^\gamma
={V\tau_p\over \sqrt{2}}
L_{1\over2}(-\kappa\partial_\mu\chi^9\partial^\mu \chi^9)
\ ,
\end{equation}
where $L_{1\over2}$ is the Laguerre function and 
$\left(\matrix{1/2\cr\gamma\cr}\right)$ are the binomial coefficients in
an expansion of $\sqrt{1+y}$.
Due to the presence of $\gamma!$, 
the expansion of $L_{1\over 2}(-\kappa\partial_\mu\chi^9\partial^\mu \chi^9)$
looks quite different to an expansion of $\sqrt{1+\kappa'\partial_\mu\chi^9
\partial^\mu \chi^9}$. However
if we take $\kappa' = 1.27\kappa$ then a plot of these two functions
looks almost identical to figure one (only with a different scale for
the vertical axis) and hence in this case there is
again a remarkable agreement.

On the other hand we may try to deduce the form of the uncompactified
non-BPS action by requiring that the effective tachyon action 
\refs{action} for a 
marginal kink $T=\chi^9(x^\mu)\ {\rm sin}\left({x^9/\sqrt{2\alpha'}}
\right)$  on   ${\bf T}^4/{\bf Z}_2$ 
reproduces the action \refs{stableaction}. This leads to the 
relation
\begin{eqnarray}
{V\over \pi}\sum_{\alpha, \beta}\sum_{k=0}^\beta  
c_{\alpha\beta}&\left({1\over 2\alpha'}\right)^{\beta-k}&
\left(\matrix{\beta\cr k\cr}\right)
B(\alpha+k+\frac{1}{2},\beta-k+\frac{1}{2})(\chi^2_9)^{\alpha+\beta-k}
(\partial_\mu\chi^9\partial^\mu\chi^9)^{k}\cr
&=&\mu\sum_{\beta=0}^\infty
\left(\matrix{\frac{1}{2}\cr \beta\cr}\right)
(\kappa'\partial_\mu\chi^9\partial^\mu\chi^9)^\beta\ ,
\label{redux}
\end{eqnarray}
where $B(x,y)$ 
is the Euler Beta-function.
First we read off the values of the coefficients
for the terms involving $\partial_\mu\chi^9$ with no undifferentiated 
$\chi^9$'s
\begin{eqnarray}
c_{0\beta} = {\mu\over V} \kappa'^\beta
\left(\matrix{\frac{1}{2}\cr \beta\cr}\right)
{\pi\over B(\beta+\frac{1}{2},\frac{1}{2})}\ .
\end{eqnarray}
Next we must insure that all other terms on the left hand side of 
\refs{redux} vanish.
This leads to the relations, for all $\beta \ge \alpha >0$,
\begin{equation}
\sum_{\gamma=\beta-\alpha}^{\beta}  
c_{\beta-\gamma\ \gamma}
\left({1\over 2\alpha'}\right)^{\gamma+\alpha-\beta}
\left(\matrix{\gamma\cr \beta-\alpha\cr}\right)
B(2\beta-\alpha-\gamma+\frac{1}{2},\gamma+\alpha-\beta+\frac{1}{2})=0\ .
\label{recrel}
\end{equation}

As before, it 
is helpful to absorb the parameters $\mu,\kappa',\alpha'$ and $V$
into the coefficients and perform some algebraic simplifications.
In particular if we let
\begin{equation}
c_{\alpha\beta} = {\mu\kappa'^{\beta}\over V}
\left({\kappa'\over 2\alpha'}\right)^\alpha {\tilde c}_{\alpha\beta}\
\end{equation}
then we find that these new coefficients are rational numbers with
\begin{equation}
{\tilde c}_{0\beta} = {(-1)^{1+\beta}\over 2\beta -1}\ ,
\label{c0}
\end{equation}
and the constraint 
\refs{recrel} can be interpreted as a recursion relation
\begin{equation}
{\tilde c}_{\alpha\beta} = -\sum_{\gamma=0}^{\alpha-1}
{\tilde c}_{\gamma\ \alpha+\beta-\gamma}
{\left(\matrix{\alpha+\beta-\gamma\cr \beta\cr}\right)
\left(\matrix{\alpha+\beta\cr \beta+\gamma\cr}\right)
\over\left(\matrix{2\alpha+2\beta\cr 2\beta+2\gamma\cr}\right)}\ .
\end{equation}
Note that the right hand side of this relation only involves knowing the
${\tilde c}$-coefficients whose first index less than $\alpha$.
On the other hand, the ${\tilde c}_{0\beta}$ are given by \refs{c0}, so 
that all coefficients ${\tilde c}_{\alpha\beta}$ 
are uniquely determined using this recursion relation. Although 
we have been unable to obtain a closed form for these coefficients,
they can, in principle, be computed to arbitrary high orders.
For example  we find, for $\alpha=1,2,3,4$,
\begin{eqnarray}
{\tilde c}_{1\beta} &=& (-1)^{1+\beta}{(\beta+1)\over (2\beta+1)^2}\ ,\cr
{\tilde c}_{2\beta} &=& {(-1)^{\beta}\over2}{(\beta+1)(\beta+2)(2\beta+7)
\over (2\beta+1)(2\beta+3)^3}\ ,\cr
{\tilde c}_{3\beta} &=& {(-1)^{\beta+1}\over2}
{(\beta+1)(\beta+2)(\beta+3)(101+44\beta+4\beta^2)
\over (2\beta+1)(2\beta+3)(2\beta+5)^4}\ ,\\
{\tilde c}_{4\beta} &=& {(-1)^{\beta}\over 8}
{(\beta+1)(\beta+2)(\beta+3)(\beta+4)(80\beta^4+2000\beta^3
+15632\beta^2+48364\beta+50759)
\over (2\beta+1)(2\beta+3)(2\beta+5)^2(2\beta+7)^5}\nonumber .\label{c12}
\end{eqnarray}
Thus we see that by
requiring  that the tachyon action reproduces 
the correct effective action for the stabilised brane on a orbifold, 
we completely fix its form.

Let now try to gain an idea as to how the resulting effective action looks.
First we can calculate a power series expansion of the tachyon 
potential
\begin{eqnarray}
V(T) &=& \sum_{\alpha=0}^\infty c_{\alpha 0}T^{2\alpha}\cr
&=& {\mu\over V}\left(1 - {\kappa' T^2\over 2\alpha'}
+{7\over 27}\left({\kappa' T^2\over 2\alpha'}\right)^2+\ldots 
\right)\ .
\end{eqnarray}
Clearly the potential is not given by $e^{-{\kappa'\over 2\alpha'}T^2}$.
This is due to neglecting the higher derivative terms. 
To fourth  order in $T$ we see that the potential is bounded from below
with a minima at $\kappa' T^2/2\alpha' =27/14 \sim 1.9 $ and 
the ratio of the energies  of the true and false vacua 
is $1/28\sim 0.04$. Unfortunately by going to higher orders in
$T^2$ one finds that the power expansion diverges for 
$\kappa' T^2/2\alpha' \sim 1$. In figure two we have plotted
the potential and the coefficient of the
$\kappa'\partial_mT\partial^m T$ 
and  $(\kappa'\partial_mT\partial^m T)^2$ terms 
as a function of $\sqrt{{\kappa'\over 2\alpha'}}T$ up to
order $T^{40}$. The series is non-convergent beyond the region plotted. 
In addition, $V(T)$ is
strictly decreasing and positive in this range, so that any 
global minima are beyond the reach of this expansion. However,
it is certainly consistent that this potential has
a minimum value of zero and that $V(T)$ is non-analytic there.

On the other hand we
can consider and expansion in powers of $T^2$ but to all orders in
$\partial_mT\partial^m T$. From ${c}_{0\beta}$ and 
${c}_{1\beta}$  we can identify
\begin{eqnarray}
{\cal L} = {\mu\over V}\Big\{&1&
+\sqrt{\kappa'\partial_m T\partial^m T}
{\rm arctan}(\sqrt{\kappa'\partial_m T\partial^m T})\cr
&-&{\kappa' T^2\over 2\alpha'}\left[
{{\rm arctan}(\sqrt{\kappa'\partial_m T\partial^m T})
\over 2 \sqrt{\kappa'\partial_m T\partial^m T}}+
{{\rm diarctan}(\sqrt{\kappa'\partial_m T\partial^m T})
\over 2 \sqrt{\kappa'\partial_m T\partial^m T} }
\right]+\ldots\Big\}\ ,
\label{weirdo}
\end{eqnarray}
where the ellipsis denotes terms with higher powers of $T^2$. In 
\refs{weirdo} ${\rm diarctan}(x)\equiv \frac{1}{2i}({\rm dilog}(1+ix)-{\rm
  dilog}(1+ix))$ where
\begin{equation}
{\rm dilog}(x) \equiv \int_1^x {{\rm ln}(t)\over 1-t}dt\ .
\end{equation}
We note that at $T=0$ and 
for large $\kappa'\partial_mT\partial^mT$ the kinetic
terms behave as ${\pi\over 2}\sqrt{\kappa'\partial_mT\partial^mT}$,
i.e. similar to the actions discussed in the previous section. Indeed if we 
choose $\kappa'=4/\pi$ then a plot of $1+\sqrt{4y/\pi}{\rm arctan}
\sqrt{4y/\pi}$ vs $y$ looks identical to the plots in figure one. 
For this choice of $\kappa'$ one finds that there is a reasonable, but
not striking,
agreement between the potential  found in this section 
(i.e. the potential plotted in 
figure two)  and that found in BSFT, i.e.  $e^{-{2\over\pi}T^2}$.

To complete the analysis we would like to find the kink solution 
and compute its energy. We will only consider the fourth order
approximation for the potential but we must also 
decide which of the kinetic terms to keep. We have considered
the following four possibilities: 
(1) we keep all terms with no more
than four powers of $T$ and/or $T'$ 
(i.e. all terms with $\alpha+\beta \le 2$), 
(2) we keep only kinetic terms quadratic in $T'$ 
but with coefficients up to order
$T^4$, (3)  we keep only kinetic terms quadratic in $T'$ but with coefficients 
up to order $T^2$, (4)  we keep only the simplest kinetic term 
$c_{01}(T')^2$.

In the first case we find that there are no solitons. More precisely,
in the region near $T=0$, the kink equation becomes complex.
In the other cases a kink soliton can be found. 
Note that  for a non-BPS D$p$-brane, 
$c_{00}$ is its tension and hence we learn that $c_{00}=\mu/V
=\sqrt{2}\tau_p$, where $\tau_p$ is the tension of a BPS D$p$-brane.
In the cases (2), (3) and (4) we find the kink energy is
$E=7.50\sqrt{\alpha'}\tau_p$, $E=7.56\sqrt{\alpha'}\tau_p$ and 
$E=7.30\sqrt{\alpha'}\tau_p$  respectively. 
On the other hand we wish to identify the kink with a BPS-D$(p-1)$-brane
in which case  its energy should be 
$2\pi\sqrt{\alpha'} \tau_p  \sim 6.28\sqrt{\alpha'}\tau_p$.  
Thus the kinks we find are
roughly $20\%$ too heavy.

\section{Conclusion}

In this note we have constructed a first derivative effective action 
for the tachyon field on unstable branes in string theory by 
probing the general ansatz with marginal deformations. While we 
still need to neglect higher derivative terms to fix the tachyon action 
completely, we expect our action to 
be a good approximation of the dynamics around the marginal kink. 
However, we also find remarkable agreement with the BSFT action 
which is constructed from a  very different, relevant tachyon profile. 
This may be an indication that higher derivative terms have little effect 
on marginal deformations, although we do not have a BPS-argument at hand 
to substantiate such a claim. Furthermore the actions discussed
here reproduce the correct BI-action for the centre of mass motion. 

For unstable branes our methods do not lead to BI-type actions 
for the tachyon as  suggested in 
\cite{Garousi,0003221}, although a $\sqrt{\partial_m T\partial^m T}$ form
appears to be generic at large momenta. 
Of course one might expect that there is a non-trivial field redefinition
between actions discussed from the point of view of the string S-matrix
and those discussed here or in BSFT. One can easily verify
that  field redefintions of the tachyon
$T\rightarrow f(T)$ cannot map the actions discussed we have discussed into 
those of  \cite{Garousi,0003221}, however we have not ruled out 
field redefinitions of the form
$T\rightarrow f(T,\partial_mT\partial^mT)$.

\section*{Acknowledgements}

We are grateful to Matthias Gaberdiel and Gerard Watts for helpful discussions.
N.D.L. has been  supported by a 
PPARC Advanced Fellowship,
the PPARC grant PA/G/S/1998/00613 and would like to thank the
University of Pennsylvania where part of this work was completed. 
I.S. was supported by DFG-SPP 1096 f\"ur Stringtheorie.


\iffigs
\newpage

\section*{Figures}

\EPSFIGURE{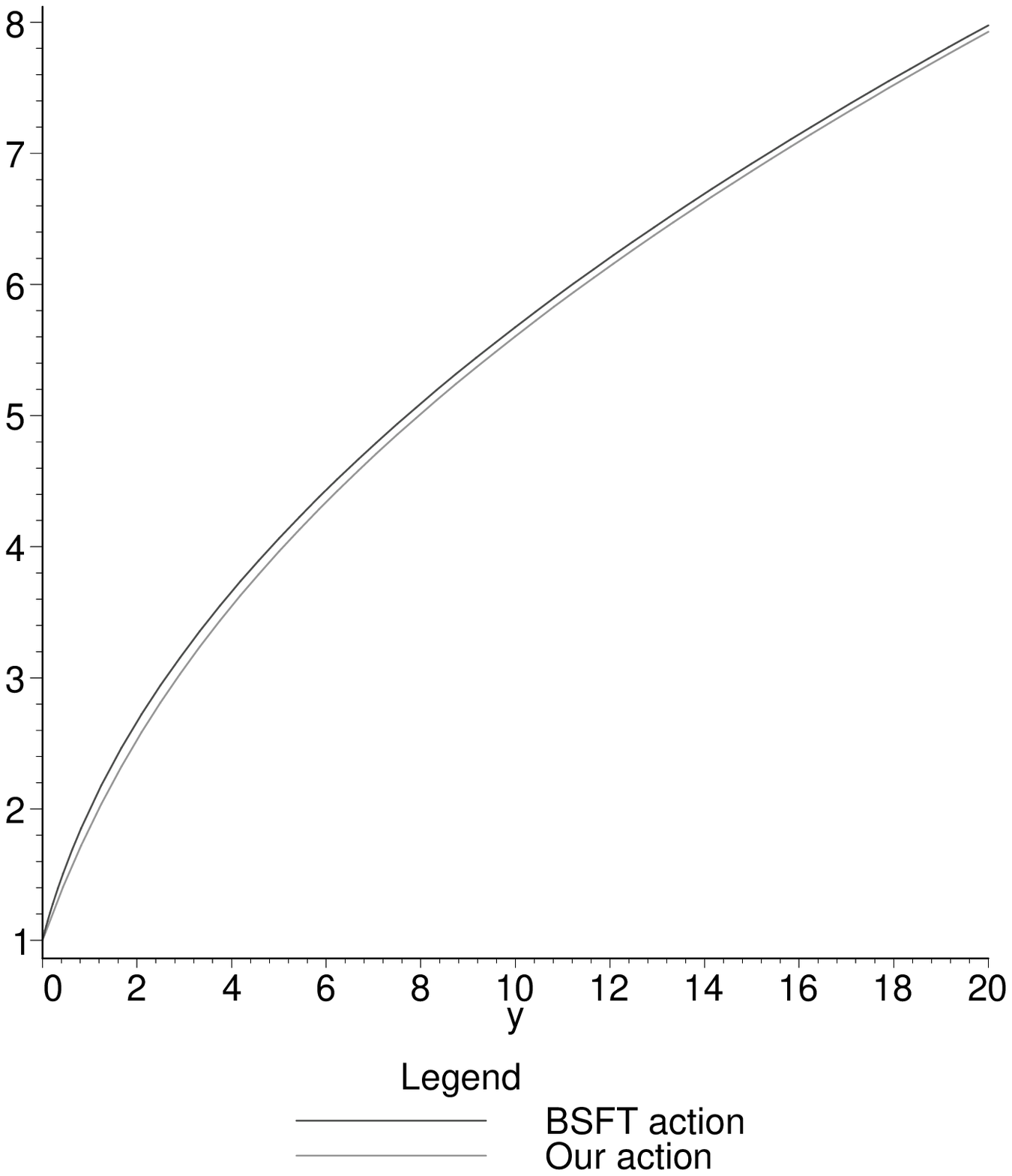}{The kinetic terms, i.e. ${\cal{L}}$ as a function of 
$y=\partial_m T\partial^m T$ at $T=0$ found in BSFT \refs{bsftaction} 
and in our action 
\refs{kinkyaction}.} 
\newpage

\EPSFIGURE{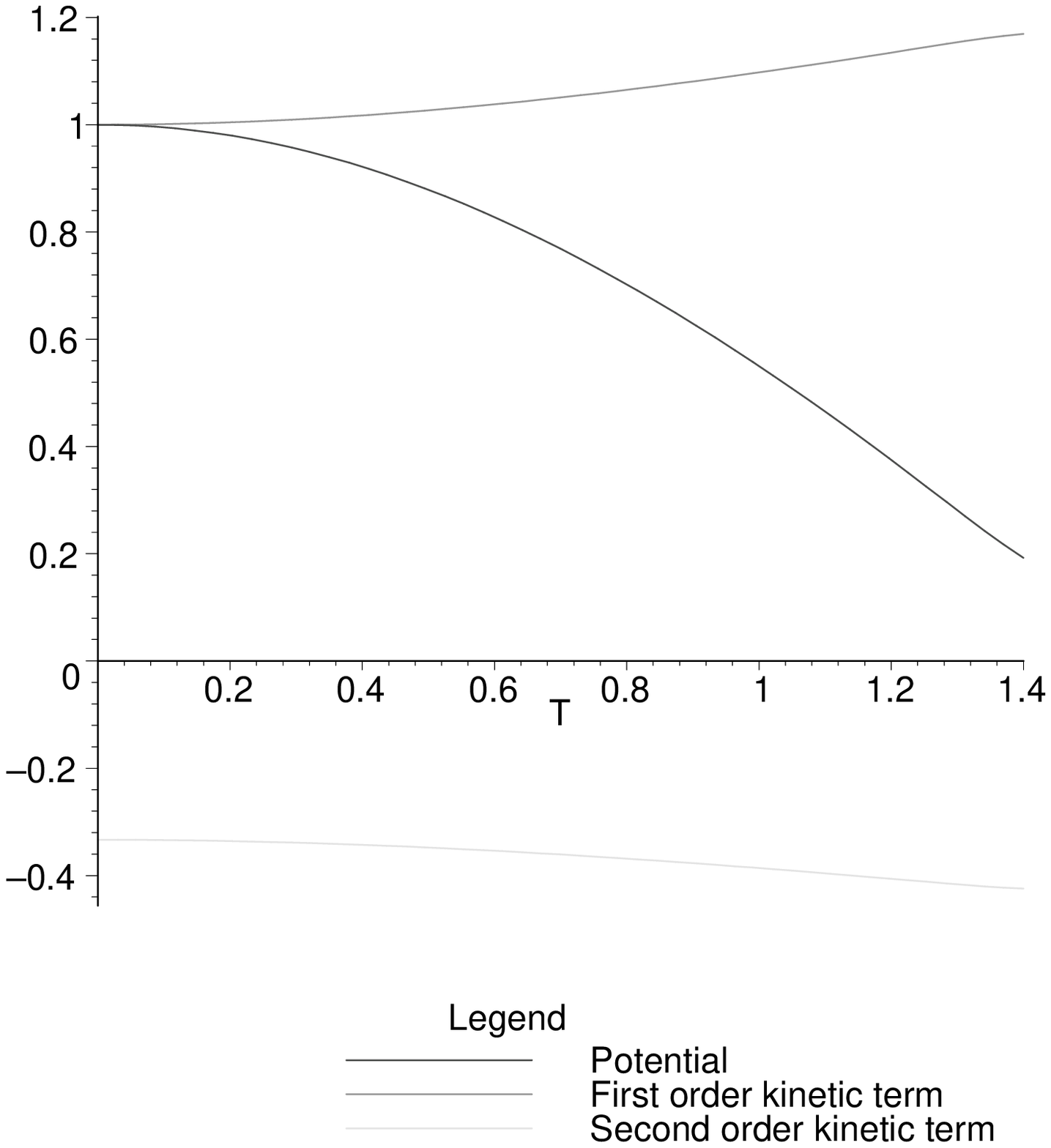}{The potential and 
coefficients of the $\kappa'\partial_mT\partial^m T$ 
$(\kappa'\partial_mT\partial^m T)^2$  terms obtained
in section three.}

\else

\message{No figures will be included. See TeX file for more
information.}
\fi

\end{document}